\documentclass[1p,times]{elsarticle}

\usepackage{graphicx}
\usepackage{subfigure}
\usepackage{amssymb}
\usepackage{amsmath}
\usepackage{epstopdf}
\usepackage{bm}

\def\simleq{\; \raise0.3ex\hbox{$<$\kern-0.75em \raise-1.1ex\hbox{$\sim$}}\; }
\def\simgeq{\; \raise0.3ex\hbox{$>$\kern-0.75em \raise-1.1ex\hbox{$\sim$}}\; }
\newcommand{\eV}{{\rm eV}}

\newcommand{\GeV}{{\rm GeV}}
\newcommand{\MV}{{\rm MV}}
\newcommand{\TeV}{{\rm TeV}}

\newcommand{\erg}{{\rm erg}}

\newcommand{\kpc}{{\rm kpc}}

\newcommand{\cm}{{\rm cm}}
\newcommand{\km}{{\rm km}}

\newcommand{\s}{{\rm s}}
\newcommand{\yr}{{\rm yr}}
\newcommand{\kyr}{{\rm kyr}}

\begin{document}
\begin{frontmatter}

\title{Implications of the Cosmic Ray Electron Spectrum and Anisotropy measured with Fermi-LAT}

\author[got]{Giuseppe Di Bernardo}
\ead{giuseppe.dibernardo@physics.gu.se}

\author[sis]{Carmelo Evoli} 
\ead{evoli@sissa.it}

\author[dip,inf]{Daniele Gaggero}
\ead{daniele.gaggero@pi.infn.it}

\author[inf,dip]{Dario Grasso  \corref{cor}}
\ead{dario.grasso@pi.infn.it}

\author[des]{Luca Maccione}
\ead{luca.maccione@desy.de}

\author[ba]{Mario Nicola Mazziotta}
\ead{Marionicola.Mazziotta@ba.infn.it}

\noindent
\cortext[cor]{Corresponding author}

\address[got]{Department of Physics, University of Gothenburg, SE-412 96, Gothenburg, Sweden}
\address[dip]{Dipartimento di Fisica, Universit\`a di Pisa, Largo Bruno Pontecorvo 3, 56127 Pisa, Italy}
\address[inf]{INFN, Sezione di Pisa, Largo Bruno Pontecorvo 3, 56127 Pisa, Italy}
\address[sis]{SISSA, via Bonomea 265, 34136 Trieste, Italy}
\address[des]{Deutsches Elektronen-Synchrotron, Notkestra{\ss}e 85, 22607
Hamburg, Germany}
\address[ba]{Istituto Nazionale di Fisica Nucleare, Sezione di Bari, 70126 Bari, Italy}


\begin{abstract}
The Fermi Large Area Telescope (LAT) collaboration recently released the updated results of the measurement of the cosmic ray electron (CRE) spectrum and published its first constraints on the CRE anisotropy. With respect to the previous Fermi-LAT results, the CRE spectrum measurement was extended down from 20 to 7 GeV, thus providing a better lever arm to discriminate theoretical models. Here we show that the new data strengthen the evidence for the presence of two distinct electron and positron spectral components. Furthermore, we show that under such hypothesis most relevant CRE and positron data sets are remarkably well reproduced. Consistent fits of cosmic-ray nuclei and antiproton data, which are crucial to validate the adopted propagation setup(s) and to fix the solar modulation potential, are obtained for the Kraichnan and plain-diffusion propagation setups, while the Kolmogorov one is disfavored. We then confirm that nearby pulsars are viable source candidates of the required $e^\pm$ extra-component.  In that case, we show that the predicted CRE anisotropy is compatible with Fermi-LAT constraints and that a positive detection should be at hand of that observatory. Models assuming that only nearby supernova remnants contribute to the high energy tail of the observed CRE spectrum are in contrast with anisotropy limits. 
\end{abstract}



\end{frontmatter}

\section{Introduction}

Last year, the Fermi-LAT Collaboration published the electron + positron spectrum in the energy range between 20 GeV and 1 TeV, measured during the first six months of the Fermi mission \cite{Abdo:2009zk}.
That result came in the middle of a hot debate which arose as ATIC \cite{atic} and PAMELA \cite{Pamela_pos} collaborations reported some anomalies in the Cosmic Ray (CR) leptonic sector. 
In particular, ATIC observed a pronounced bump 
in the electron + positron spectrum at around 600 GeV, while PAMELA found the positron fraction $e^+/(e^+ + e^-)$ to increase with energy above 10 GeV. Both features are hardly compatible with the standard scenario in which CR electrons ($e^-$) are accelerated in the surrounding of Galactic supernova remnants (SNRs) and positrons are predominantly of secondary origin (i.e.~they are produced only by spallation of CR nuclei onto the interstellar medium gas). 
Below 100 GeV the Cosmic Ray Electron (CRE) spectrum observed by the Fermi-LAT agrees with the one measured by ATIC, but at higher energies it does not display the feature seen by that experiment, being compatible with a single power law with index $3.045 \pm 0.008$. The absence of the ATIC feature was confirmed by the H.E.S.S. atmospheric Cherenkov telescope \cite{hess:09}. Below 1 TeV the spectrum measured by H.E.S.S. is in agreement with Fermi-LAT's while above that energy (a range not probed by Fermi-LAT) a significant spectral steepening was reported by the former experiment.

Soon after that measurement was released, several members of the Fermi-LAT Collaboration showed -- using the {\tt GALPROP} package \cite{galprop,GALPROPweb} --  that an interpretation of the observed spectra is possible within a conventional model in which SNRs are the only primary sources of CRE \cite{CRE_interpretation1}. Below 20 GeV, however, they found the predictions of that model to be in tension with pre-Fermi measurements, in particular HEAT \cite{Barwick:1997ig}  and AMS-01 \cite{AMS01} data. Furthermore,  the positron fraction measured by PAMELA was not reproduced within that framework neither above (where its observed slope is opposite to the theoretically expected one) nor below $\sim 10~\GeV$. 
An alternative scenario was therefore proposed in the same paper \cite{CRE_interpretation1}, which invoked the presence of an extra component of electrons and positrons. In order to reproduce the observed CRE spectrum and positron fraction, the injection spectrum of such extra component has to be harder than the one expected to be accelerated by SNRs and must display a cutoff at around 1 TeV. In that case it was shown that both Fermi-LAT and PAMELA measurements above 20 GeV could be explained consistently and the disagreement of the model with pre-Fermi data below 20 GeV was reduced (the remaining discrepancy was guessed to be a consequence of solar modulation). 

Following the suggestion of some pioneering works \cite{Shen,Aharonian:1995zz} (see also the more recent \cite{Hooper:2008kg}), in \cite{CRE_interpretation1} it was argued that such extra-component can naturally be produced by middle-aged pulsars in the vicinity of the Solar System. An alternative scenario based on the annihilation of dark matter particles with mass around $1~\TeV/c^2$  was also discussed in the same paper and shown to be compatible with Fermi-LAT, H.E.S.S. and PAMELA data (see also \cite{Bergstrom:2009fa} and a number of following papers, e.g.~\cite{Cirelli:2008jk,Cholis:2008hb}). Several  counter-arguments, however, have been risen which make this interpretation disfavored if not ruled out (see e.g.~\cite{Ackermann:2010rg}). 
Other possible origins of the required electron and positron extra-component have also been proposed, e.g., enhanced secondary production in standard SNRs \cite{Blasi:2009hv} or an inhomogeneous distribution of source in the Solar System neighborhood \cite{Shaviv:2009bu}. 


Recently, the Fermi-LAT Collaboration released a new measurement of the CRE spectrum based on one year data. The observed spectrum extends down to 7 GeV \cite{Ackermann:2010ij} and is confirmed to be compatible with a single power-law with a spectral index $3.08 \pm 0.05$, only slightly softer than the one published in \cite{Abdo:2009zk}. 
Hints of a deviation from a pure power-law behavior between 20-100 GeV and at a few hundreds GeV,  which were found in the six month data, are still present in the updated spectrum. The most relevant new piece of information comes from the data in the 7 - 20 GeV energy range. Interestingly, the updated Fermi-LAT spectrum agrees with HEAT \cite{Barwick:1997ig} and AMS-01 \cite{AMS01} data. As we will show these findings rise serious troubles to single component models and call for some revisions also for the two-component model discussed in \cite{CRE_interpretation1}. 
 %

Such problems become even more challenging if one considers   
complementary CR data sets, in particular the positron-to-electron fraction and the antiproton spectrum measured by several other experiments. Interestingly, even at low energy several discrepancies can be found between the data and the predictions of numerical CR diffusion models,
as {\tt GALPROP} or {\tt DRAGON} ~\footnote{Code available at http://www.desy.de/$\sim$maccione/DRAGON/} \cite{Evoli:2008dv}, if a conventional Kolmogorov propagation setup is adopted. Indeed, the positron fraction measured by PAMELA is significantly lower than the numerical estimates obtained within this setup below 10 GeV (see e.g.~\cite{Pamela_pos,CRE_interpretation1}). A long-standing problem is represented by the antiproton deficit which was found by the {\tt GALPROP} team with respect to several experimental data \cite{Moskalenko:2001ya}.  Recently it became more serious with the release of PAMELA antiproton high statistics data \cite{Adriani:2008zq,pamela:2010rc} (see \cite{DiBernardo:2009ku}).

A possible solution of some of those anomalies has been recently proposed in terms of charge-dependent solar modulation \cite{Gast}. To explain  
Fermi-LAT data, however, this scenario requires to assume quite extreme conditions during the latest solar activity phase. Furthermore, its predictions disagree with the preliminary measurements of the absolute electron ($e^-$) spectrum performed by PAMELA below $\sim 7~\GeV$  \cite{Adriani_talk}. It is important to mention here that Fermi-LAT and PAMELA have been taking data during the same solar cycle, which allows to reduce the uncertainties related to solar modulation.

Here we aim at obtaining a consistent description of all the observations mentioned above following a different approach: we work in a simple force field charge independent modulation framework but we consider other propagation setups besides the Kolmogorov one adopted in \cite{CRE_interpretation1,Ackermann:2010ij}. In particular, we adopt a model based on Kraichnan-like turbulence with a lower level of reacceleration with respect to the Kolmogorov one: this model was presented in \cite{DiBernardo:2009ku} and obtained as a result of a combined maximum likelihood analysis based on B/C data (including the recent CREAM data-set \cite{CREAM}) and ${\bar p}/p$ updated measurements.  We will show that a consistent description of most CR data sets is indeed possible for models of this kind. 


This work is organized in the following way: in Sec.~\ref{sec:propagation} we will identify three reference propagation setups compatible with cosmic-ray nuclei observations;  then in Sec.~\ref{sec:single_comp} we will adopt those setups trying to build single-component CRE models and showing why they do not provide a satisfactory description of the available data; in Sec.~\ref{sec:toy_model}  we will describe how the introduction of an $e^\pm$ extra-component allow a consistent interpretation of those data; finally, in Sec.~\ref{sec:discrete_sources} we will show how pulsars can rather naturally power such extra-component, and check if the proposed scenario is compatible with upper limits on the CRE anisotropy recently released by the Fermi-LAT collaboration \cite{Ackermann:2010ip}. Sections 6 and 7 are devoted to a discussion of our findings and to concluding remarks.

\section{The propagation setups}\label{sec:propagation}

%
The CR propagation equation from a continuous distribution of sources can be written in the general form:
\begin{eqnarray}
\frac{\partial N^i}{\partial t} &-& {\bm \nabla}\cdot \left( D\,{\bm \nabla}
-\bm{v}_{c}\right)N^{i} + \frac{\partial}{\partial p} \left(\dot{p}-\frac{p}{3}\bm{\nabla}\cdot\bm{v}_{c}\right) N^i -\frac{\partial}{\partial p} p^2 D_{pp}
\frac{\partial}{\partial p} \frac{N^i}{p^2} =  \nonumber \\
&=&  Q^{i}(p,r,z) + \sum_{j>i}c\beta n_{\rm gas}(r,z)
\sigma_{ji}N^{j} -  c\beta n_{\rm gas}\sigma_{\rm in}(E_{k})N^{i}\;.
\label{eq:diffusion_equation}
 \end{eqnarray}
Here $N^i(p,r,z)$ is the number density of the $i$-th atomic species; $p$ is its momentum; $\beta$ the particle velocity in units of the speed of light $c$; $\sigma_{in}$ is the total inelastic cross section onto the inter stellar medium (ISM) gas, whose density is $n_{\rm gas}$; $\sigma_{ij}$ is the production cross-section of a nuclear species $j$ by the fragmentation of the $i$-th one; $D$ is the spatial diffusion coefficient; $\bm{v}_{c}$ is the convection velocity. The last term on the l.h.s.~of Eq.~(\ref{eq:diffusion_equation}) describes diffusive reacceleration of CRs in the turbulent galactic magnetic field. In agreement with the quasi-linear theory we assume the diffusion coefficient in momentum space $D_{pp}$ to be related to the spatial diffusion coefficient by the relationship (see e.g.~\cite{Berezinsky:book}) 
\begin{equation}
D_{pp} = \frac{4}{3 \delta (4 - \delta^2)(4 - \delta)} \frac{v_A^2~p^2}{D}\;,
\label{eq:dpp}
\end{equation}
where  $v_A$ is the Alfv\`en velocity and $\delta$ the index of the power-law dependence of the spatial diffusion coefficient on the particle rigidity
(see Eq.\ref{eq:diff_coeff}). Here we assume that diffusive reacceleration takes place in the entire diffusive halo. 

The distribution of Galactic CR sources is poorly known. The large scale source distribution which we adopt here is approximated to be cylindrically symmetric, so that the source term for every nuclear species $i$ takes the form:
\begin{equation}
Q_{i}(E_{k},r,z) =  f_S(r,z)\  q_{0,i}\ \left(\frac{\rho(E_{k})}{\rho_0}\right)^{- \gamma_{i}} \;,
\label{eq:source}
\end{equation}
For CR nuclei, we assume that $f_S(r,z)$ traces the distribution of supernova remnants and impose the normalization condition $f_{S}(r_{\odot},z_{\odot}) = 1$.  We use the same distribution as in \cite{DiBernardo:2009ku} for all models considered in this paper.  This is slightly different from that used in {\tt GALPROP} which, however, does not have significant effects on the CRE and positron spectra.
We also assume that the injection spectral index is independent of the primary nucleus and is the same as that of protons $\gamma_ {i} = \gamma_{p}$. Since SNRs are expected to accelerate CR nuclei up to at least $10^{15}~\eV$, which is well above the energies we consider here, we do not account for a possible high energy suppression of the injection spectrum. 

We neglect CR convection as we tested that under reasonable conditions its effects are negligible at the energies under consideration here and that it is not necessary to reproduce experimental data. Nuclei spallation is treated as described in \cite{GALPROPweb,Strong:98,Strong:04}. 

Here we assume the diffusion coefficient $D$ to be spatially uniform and that it only depends on the particle rigidity $\rho$ and the particle speed $\beta$ according to the following relation
\begin{equation}
\label{eq:diff_coeff}
 D(\rho) = D_0 ~\beta^\eta \left(\frac{\rho}{\rho_0}\right)^\delta \;,
 \end{equation}
with $\eta$ controlling essentially the low energy behavior of $D$. While one would expect  $\eta=1$ as the most natural dependence of diffusion on the particle speed,
it should be taken into account that diffusion may actually be inhibited at low energies due to the back-reaction of CR on the magneto-hydrodynamic waves.
 In a dedicated analysis \cite{Ptuskin:2005ax} of that effect, a low-energy increase of $D$  was found. While such a behavior cannot be represented as a simple function of $\beta$ and $\rho$,  an effective value of $\eta$ can, nevertheless, be found which allows to fit low energy data. 
Clearly, the precise value depends on the details of the model. For example in \cite{Ptuskin:2005ax} $\eta = - 3$ was found to fit the data, while the authors of \cite{Maurin:09} found $\eta = -1.3$, in both cases for models with $\delta = 0.5$ (but rather different values of other parameters). In \cite{DiBernardo:2009ku} we found that $\eta = -0.4$ allows a rather good fit of low energy nuclear data for models with low reacceleration and $\delta \simeq 0.5$. 

\begin{figure}[tbp]
  \centering
   \includegraphics[scale=0.5]{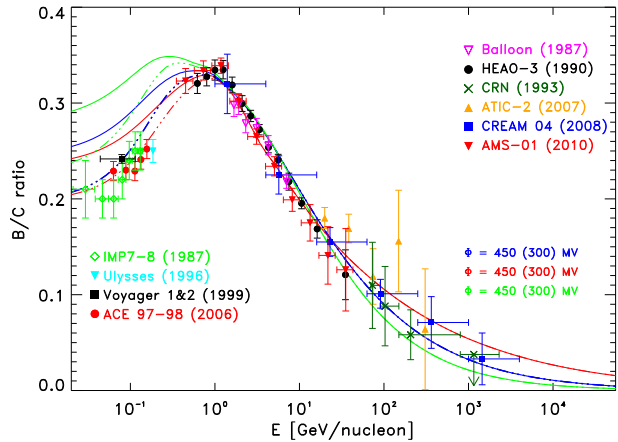}
\caption{B/C ratios, as computed with our three reference models given in Tab.~\ref{table1}, are compared with experimental data. Red lines: Kolmogorov model  (KOL);  blue lines: Kraichnan model (KRA); green lines: plain diffusion (PD).  Solid lines are modulated with $\Phi = 450~{\rm MV}$, which is appropriate for HEAO-3 and CREAM data, while triple-dotted lines have been computed for $\Phi = 300~{\rm MV}$ which is appropriate for ACE data.}
\label{fig:BoverC}
\end{figure}

%
\begin{table}[tbp]
\centering
\caption{Propagation and CR injection parameters for the three reference models considered in this paper. $D_0$ is the diffusion coefficient normalization at $3$ GV; 
$\delta$ the index of the power-law dependence of $D$ on energy; $z_h$ the half-height of the Galactic CR confinement halo; $\gamma_{p}$ the CR nuclei injection index; $v_A$ is the Alfv\`en velocity; $\eta$ is the exponent of the power-law dependence of $D$ on the particle velocity $\beta$. The source spectral break rigidity for the KOL model is 4 GV.} 
\vskip0.3cm
\begin{tabular}{|c|c|c|c|c|c|c|}
\hline
Model & $\delta$ & $D_0\  (\cm^2\s^{-1})$  &   $z_h\ (\kpc)$ & $\gamma_p$ & $v_A\ (\km~ \s^{-1})$  & $\eta$ \cr
\hline
\hline 
 KOL    & 0.33 & $5.6\times 10^{28}$ & 4  & 1.6/2.4 & $v_A = 30$ & 1  \cr
 KRA  &  0.5 &  $3.0 \times 10^{28}$ & 4  & 2.25 & $v_A = 15$  & -0.4 \cr  
 PD  &   0.60 & $2.4 \times 10^{28}$ & 4  & 2.15 & $v_A = 0$ & -0.4 \cr
\hline
\end{tabular}
\label{table1}
\end{table}%
Although the description of propagation offered by Eq.~(\ref{eq:diffusion_equation}) is already extremely simplified and in fact just an effective one, present CR data are still not accurate enough to fix the values of the main parameters controlling propagation: $D_{0}$ and $\delta$ from Eq.~(\ref{eq:diff_coeff}), $v_{A}$ from Eq.~(\ref{eq:dpp}), the height of the Galactic diffusion region $z_{h}$, and the injection index $\gamma_{p}$ appearing in Eq.~(\ref{eq:source}). Moreover, when considering data below a few GeV/n also the parameter $\eta$ (see Eq.~(\ref{eq:diff_coeff})) and the modulation due to solar activity play a significant role and must be taken into account. For these reasons, here we consider three reference propagation models defined by the parameters reported in Tab.~\ref{table1}. In making this choice, we fixed $z_{h} = 4~\kpc$.~\footnote{We verified that, as long as $1~\kpc < z_{h} < 10~\kpc$, changing $z_{h}$ almost amounts to a redefinition of the value of $D_{0}$.}  Models are defined as follows: the PD model is a plain diffusion one, in which we tried to reproduce CR data with only diffusion and spallation losses, hence setting $v_{A} = 0$; the KOL model is built assuming that the power spectrum of the galactic turbulent magnetic field is of Kolmogorov type, which fixes $\delta = 1/3$, while the KRA model assumes a Kraichnan spectrum for the galactic turbulent magnetic field, hence setting $\delta = 1/2$. With these choices being made for the physical models, the remaining free parameters were tuned to reproduce the boron to carbon ratio (see Fig.~\ref{fig:BoverC}) in the energy range $0.1 \lesssim  E \lesssim 10^3~\GeV/{\rm n}$. We take experimental data from HEAO-3 \cite{HEAO-3} and CRN \cite{CRN} satellite-based experiments and from CREAM \cite{CREAM} and ACE/CRIS \cite{ACE}.  HEAO-3 B/C data are nicely confirmed by a recent analysis of AMS-01 data \cite{AMS1_BC}.

\begin{figure}[tbp]
  \centering
   \includegraphics[scale=0.5]{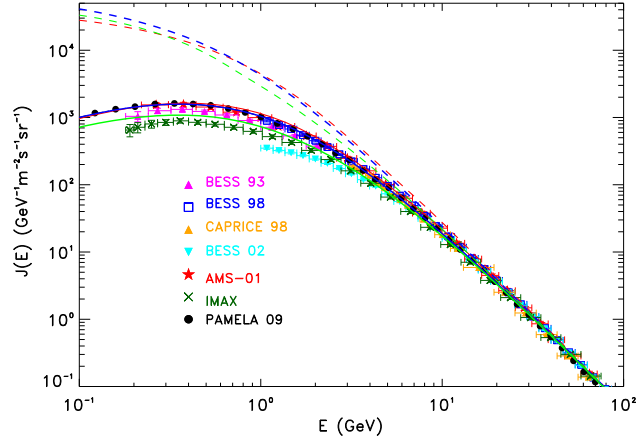}
\caption{The proton spectra calculated for our three reference models. Color notation is the same as in the previous figure.
The solar modulation potential, which was tuned to reproduce experimental during solar activity minimum, is $\Phi = 550~{\rm MV}$ for all models.
Dashed lines are the corresponding local interstellar (LIS) spectra.
}
   \label{fig:protons}
\end{figure}

We treat the solar modulation adopting the usual force-free field approach \cite{Gleeson} and tune the modulation potential $\Phi$ so as to reproduce the proton spectrum measured by PAMELA \cite{PAMELA:proton} during the same solar cycle period during which Fermi-LAT performed the CRE spectrum measurement (see Fig.~\ref{fig:protons}). 

It is clear from Tab.~\ref{table1} and from Fig.~\ref{fig:BoverC} that the steeper the energy dependence of $D$ (i.e.~the larger the value of $\delta$) the lower is the amount of reacceleration required to reasonably match the B/C data. Indeed, if $v_{A} = 0$ we need $\delta = 0.6$, while if $v_{A}=30~\km/\s$ then $\delta = 0.33$ is enough to reproduce high energy B/C data. Although the interpretation of low energy data is not straightforward, because of several poorly known effects (e.g., convection or dissipation of Alfv\`en modes) which are expected to play an increasing role with decreasing energy, we found that a proper tuning of the parameter $\eta$ is enough to provide an effective description of CR data even well below the GeV (see also \cite{DiBernardo:2009ku}). 

Although the B/C ratio is the most used quantity for the purpose of fixing CR propagation parameters, also other quantities, in particular the ${\bar p}/p$ ratio or the antiproton absolute spectrum provide complementary constraints. Indeed, ratios of nuclear species like B/C are weakly insensitive to the injection spectra of primaries, since the secondaries have almost the same energy/nucleon as the primaries. Hence we have still the freedom to adjust the injection spectra to reproduce the proton and other primary spectra. Antiprotons, on the other hand, are produced in a spectrum by the spallation of the primary protons and Helium nuclei, which depends significantly on the energy of the primary particle. Therefore, antiprotons are sensitive to the injection spectrum of primaries, and break the degeneracy. 
%
%

Constraints coming from antiproton data became quite stringent with the recent release of PAMELA data \cite{Adriani:2008zq,pamela:2010rc} and cannot be ignored.  
PAMELA proton data are also crucial to set, for each model,  the solar modulation potential.  It is interesting that all models considered here match those data for the same value of the modulation potential (see Fig.~\ref{fig:protons}) \footnote{It should be noted that differently  from the KRA and PD models, 
 the KOL model requires 
a pronounced break  in the proton injection spectrum in order to reproduce the data (see Tab.~1 and  Fig.~\ref{fig:protons}).  }.

\begin{figure}[tbp]
  \centering
    \includegraphics[scale=0.5]{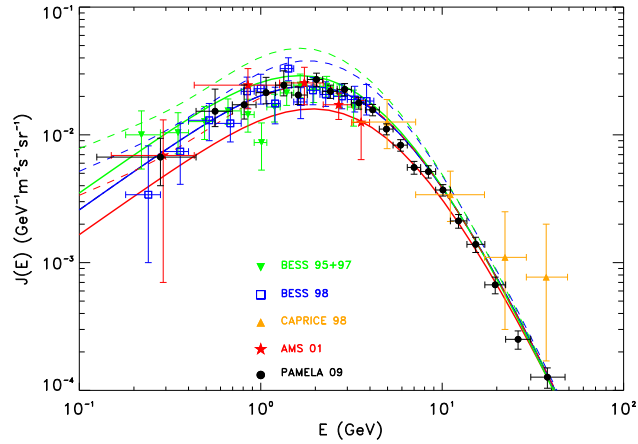}
\caption{The theoretical antiproton spectra computed for our three reference models are compared with experimental data. The color notation is the same as in the previous figures. All solid lines are modulated according to the force field scheme assuming the charge symmetric potential $\Phi = 550~{\rm MV}$
as required to reproduce the measured proton spectrum at low energies (see Fig.~\ref{fig:protons}). Dashed lines are the corresponding LIS spectra.
}
  \label{fig:antip_p}
\end{figure}

Once the modulation potential has been fixed in that way, model predictions can be compared with experimental antiproton data. 
It is clear from Fig.~\ref{fig:antip_p} that while the propagation models considered in the above are almost degenerate against the B/C data some of them, in particular the KOL model, are disfavored by the antiproton data.  This was also established on the basis of a recent combined statistical analysis of CR nuclei and antiproton data \cite{DiBernardo:2009ku}.

Overall, this seems to favor models with low reacceleration and a value of the diffusion coefficient slope close to 0.5 or higher, which is also in agreement with independent findings \cite{Maurin:2001sj}.
In the next section we will see as CR electron and positron data also lead to a similar conclusion.

%


\section{Modeling the CRE spectrum with a single Galactic component}
\label{sec:single_comp}


We start our analysis by trying to interpret CR electron and positron data with single-component models. 

We evaluated these model with the {\tt DRAGON} numerical package \cite{DiBernardo:2009ku}. We also verified that our results are reproduced by the {\tt GALPROP} package \cite{galprop,GALPROPweb} under the same physical conditions.
 We consider the three diffusion setups, KOL, KRA and PD discussed in the previous section and for each of them we tune the free parameters involved in the calculation of the CRE spectrum against Fermi-LAT data in the energy range $20 - 100~\GeV$, where both statistics and systematics errors are smallest. The choice to restrict our model fitting to data in this energy range is also strongly motivated by theoretical arguments. 
Indeed, above 20 GeV the uncertainties on the CRE spectrum due to solar modulation  are negligible; below 100 GeV the CRE spectrum is hardly affected by the uncertainties due to the stochastic spatial and temporal distribution of CRE local sources which instead become relevant in the TeV energy range \cite{Pohl:1998ug, CRE_interpretation1}.  

\begin{figure}[tbp]
  \centering
   \subfigure[]
  {
   \includegraphics[scale=0.4]{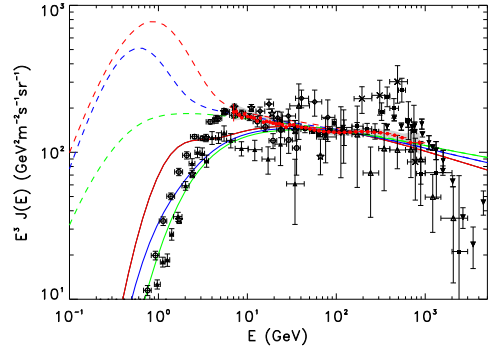}
   \label{fig:bump}
   }
     \subfigure[]
  {
   \includegraphics[scale=0.4]{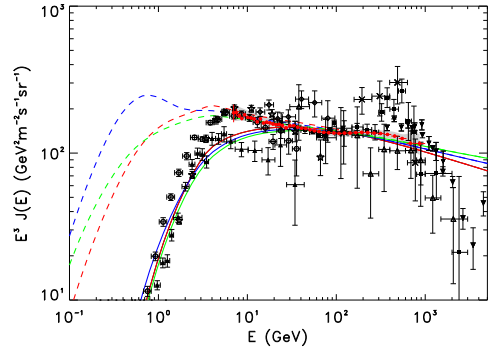}
   \label{fig:bump_break}
   } 
\caption{The $e^- + e^+$ spectra computed for single component models with unique injection slopes $\gamma = 2.45,\  2.37,\  2.32$ 
(panel a) and broken power-law $\gamma = 1.6/2.4;\ 2.0/2.37;\ 2.32/2.32$ below/above 2 GeV for the KOL (red), KRA (blue) and PD (green) diffusion setups respectively (panel b).  Fermi-LAT data are represents by the red points. See Fig.~5a,c,e for the other experiment's symbols. Modulated lines (solid curves) have been computed in the force field framework with $\Phi = 550\, {\rm MV}$; dashed lines are the corresponding local interstellar (LIS) spectra. 
}
\end{figure}

We find that above 20 GeV the injection spectral indexes  $\gamma = 2.45,\  2.37,\  2.32$, respectively for the KOL, KRA and PD models, are adequate to reproduce Fermi-LAT data (see Fig.~\ref{fig:bump}). 

Problems appear, however, when considering lower energy data.  A discrepancy between low energy pre-Fermi data and the prediction of single component models was already noticed in \cite{CRE_interpretation1} and tentatively ascribed to solar modulation or to systematics experimental errors.  That interpretation, however, does not hold against the new Fermi-LAT data between 7 and 20 GeV which agree with AMS-01 and HEAT data and are clearly incompatible with the results of the above models. This is the case both if single power-law source spectra are adopted and if a spectral break is introduced, as shown in Fig.s~\ref{fig:bump} and \ref{fig:bump_break} respectively.
 
It is worth noticing that reacceleration models need spectral breaks to correct for the anomalous behavior which would otherwise arise in the propagated LIS spectrum. 
The reason of such a behavior can be traced back to the combined effect of reacceleration and energy losses.
 Reacceleration shifts electrons from the low-energy to the high-energy region of the spectrum, while energy losses take electrons from high to low energy. The two effects have comparable strength in the GeV region and give rise to pronounced bumps in the unmodulated spectra shown in Fig.~\ref{fig:bump} if a single power-law is assumed. 
Clearly, this feature is more evident in models with strong reacceleration (as in the KOL model), and must be treated by introducing a sharp, and hardly justifiable, break in the injection. On the other hand, the KRA model, in which only moderate reacceleration is present, requires a smoother break.  No break at all is required for the PD model. 

As it is evident from Fig.s~\ref{fig:bump} and \ref{fig:bump_break}, a very small modulation potential (smaller than $\sim 100$ MV) would allow to reproduce low energy Fermi-LAT data.  Two major problems, however, arise against this option: {\it a}) such a low modulation potential is not compatible with proton and nuclei data;  {\it b}) it does not fit AMS-01 data which were taken during a solar minimum phase like Fermi-LAT ones were.
 
In principle, both problems could be circumvented in a charge asymmetric modulation scenario, in which solar modulation is expected to act differently on particles of opposite electric charge during periods of opposite polarity of the Solar magnetic field.  This is the case for AMS-01 and Fermi data taking periods. Interestingly, the antiproton and positron fraction measured by PAMELA and other experiments can be consistently reproduced in that framework \cite{Gast}. This possibility, however, seems to be at odds with recent (still preliminary, see \cite{Adriani_talk}) measurements of the electron $e^-$ spectrum by PAMELA, which agree both with Fermi-LAT, between 7 and 100 GeV, and with AMS-01 below 10 GeV.  
Indeed, even adopting a vanishing potential for the $e^-$ (which is a quite extreme assumption), such models fail to match PAMELA $e^-$ data 
down to a few GeV.  Therefore, since PAMELA and Fermi-LAT data were taken during the same solar cycle phase, it is evident that the discrepancy between the prediction of single component models and low energy data cannot be ascribed to charge asymmetric solar modulation.  

\begin{figure}[tbp]
  \centering
   \subfigure[]
  {
   \includegraphics[scale=0.35]{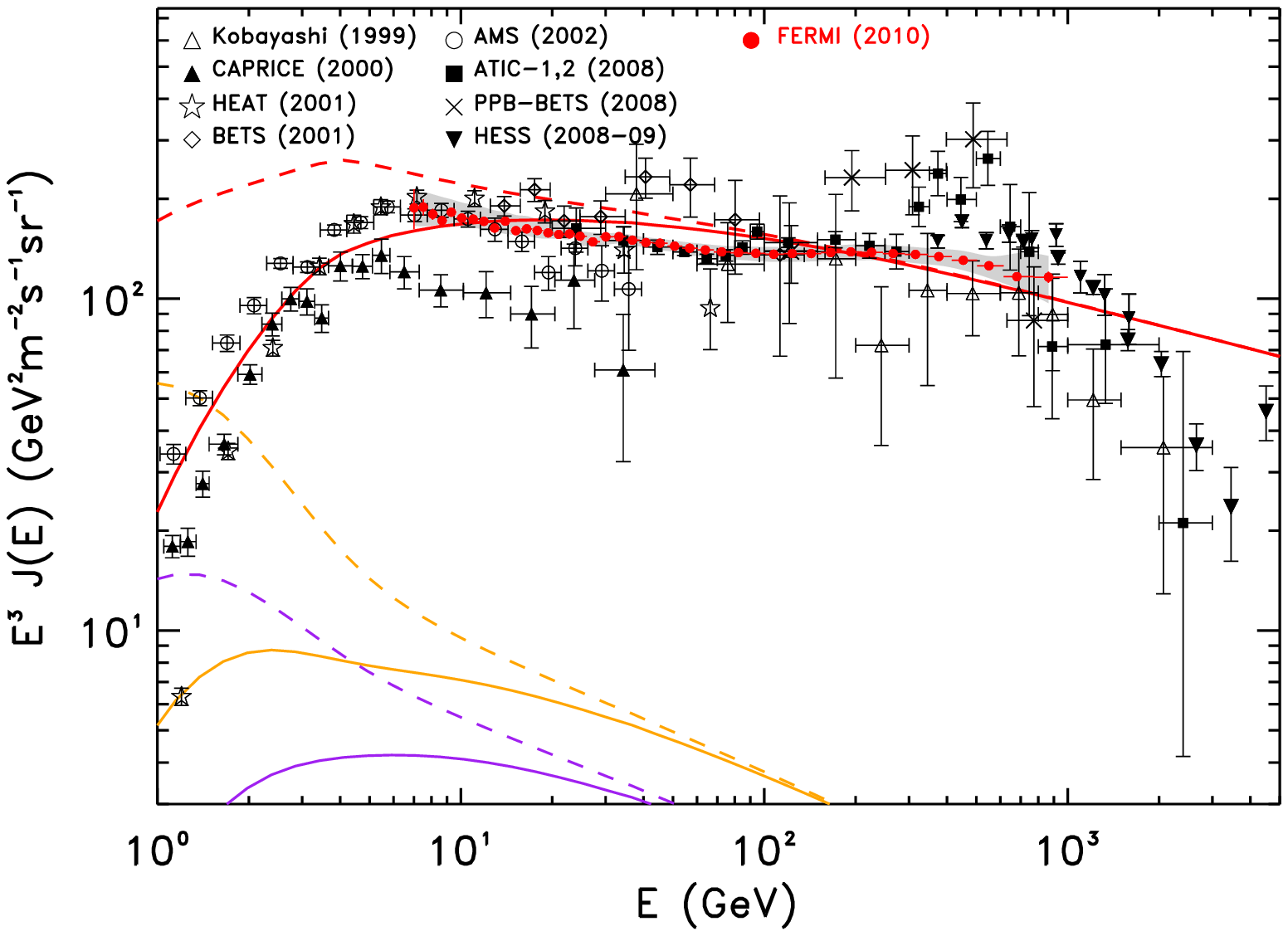}
   \label{fig:KOL_singlecomp_best}
   }
     \subfigure[]
  {
   \includegraphics[scale=0.35]{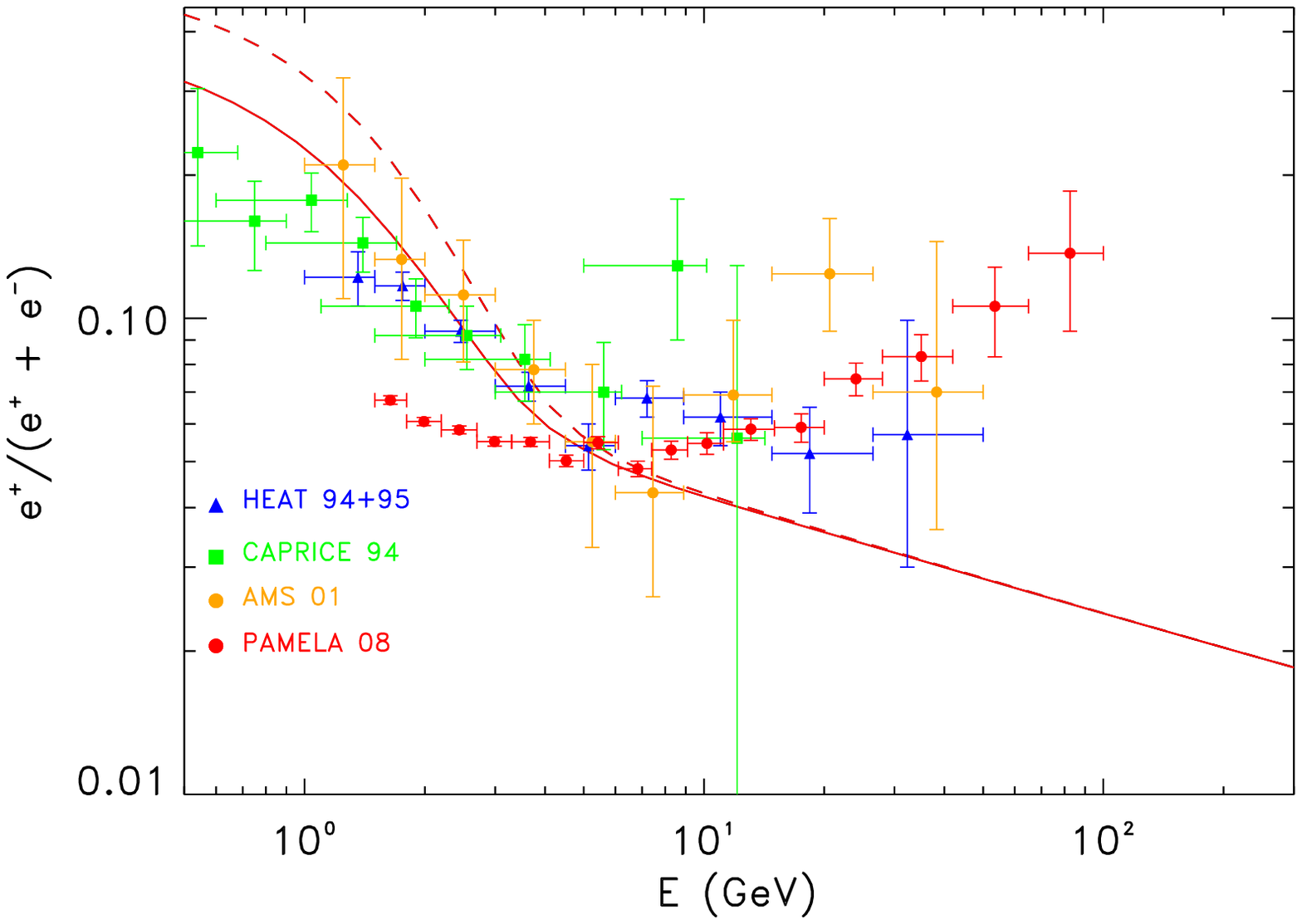}
   \label{fig:pos_KOL_singlecomp_best}
   }
    \subfigure[]
  {
 \includegraphics[scale=0.35]{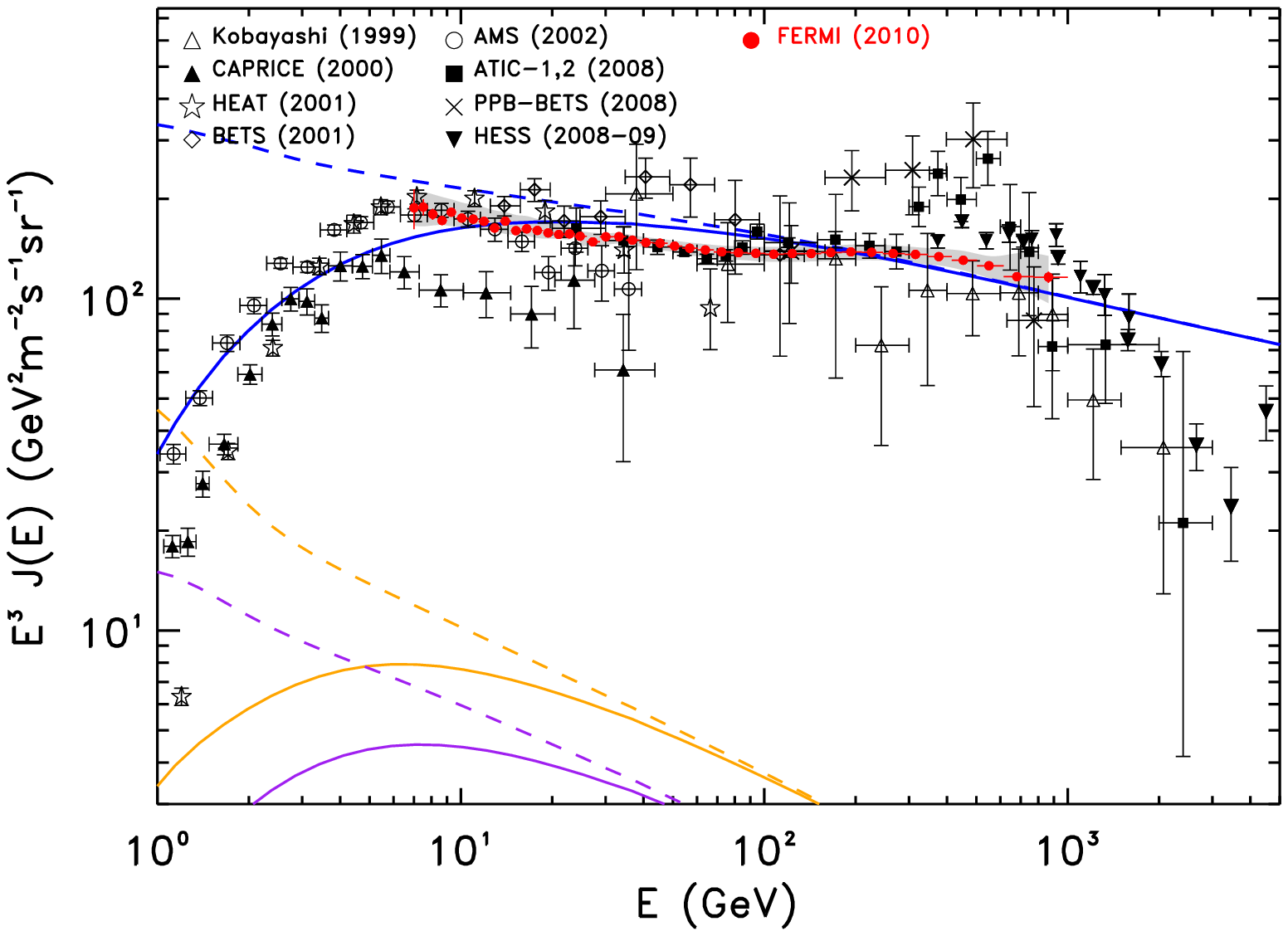}
   \label{fig:KRA_singlecomp_best}
   }
      \subfigure[]
  {
 \includegraphics[scale=0.35]{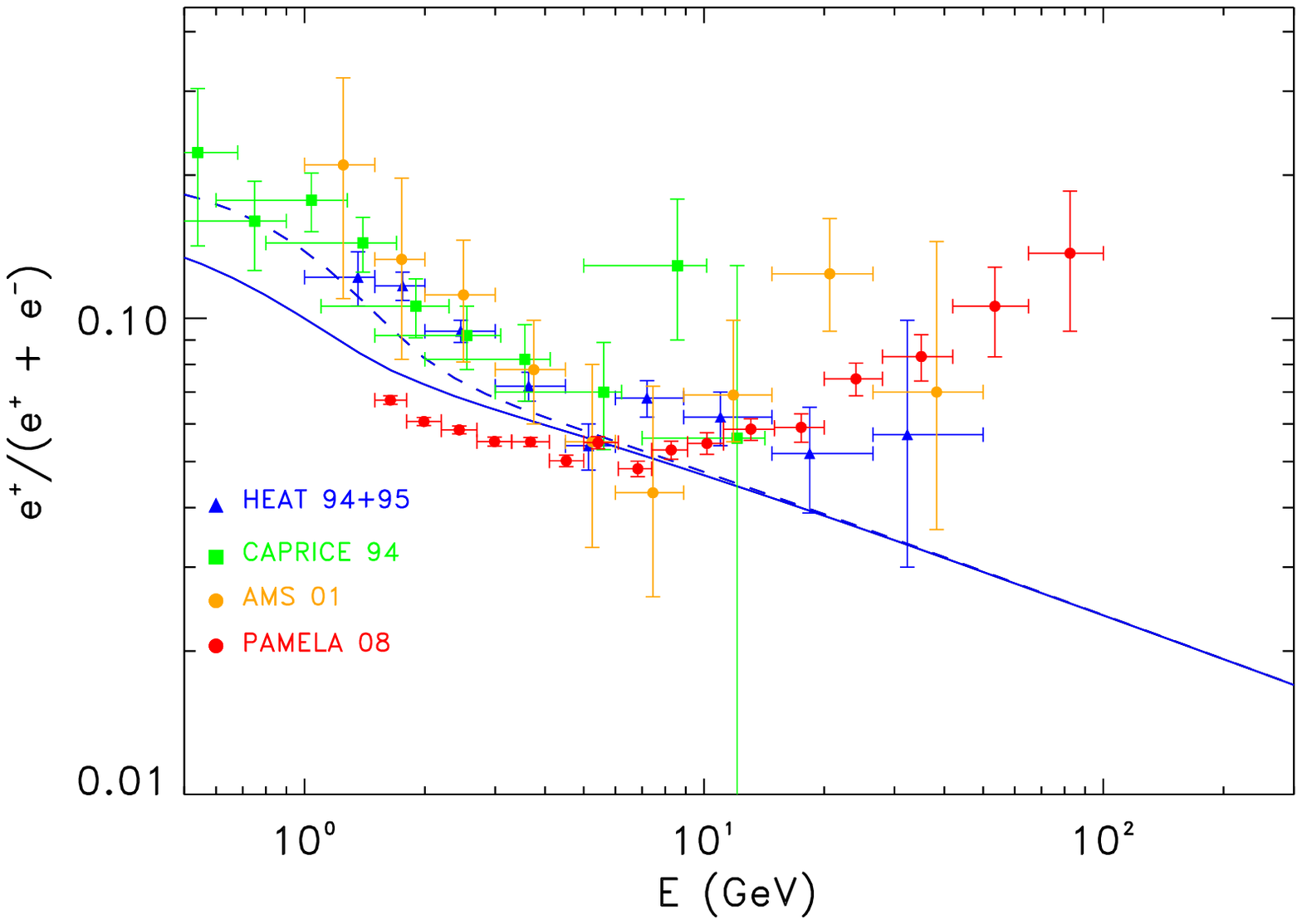}
   \label{fig:pos_KRA_singlecomp_best}
   } 
       \subfigure[]
  {
 \includegraphics[scale=0.35]{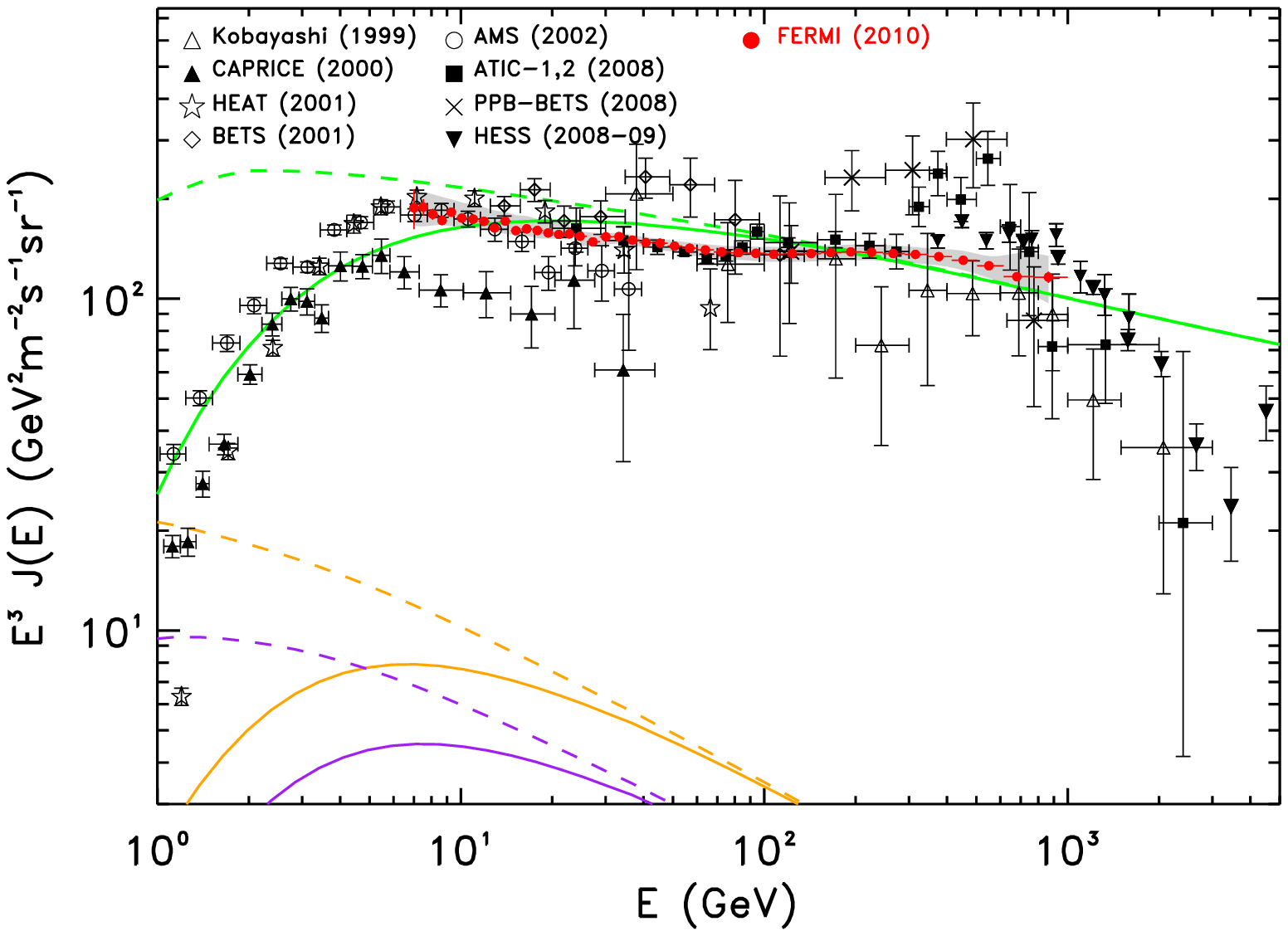}
   \label{fig:PD_singlecomp_best}
   }
      \subfigure[]
  {
 \includegraphics[scale=0.35]{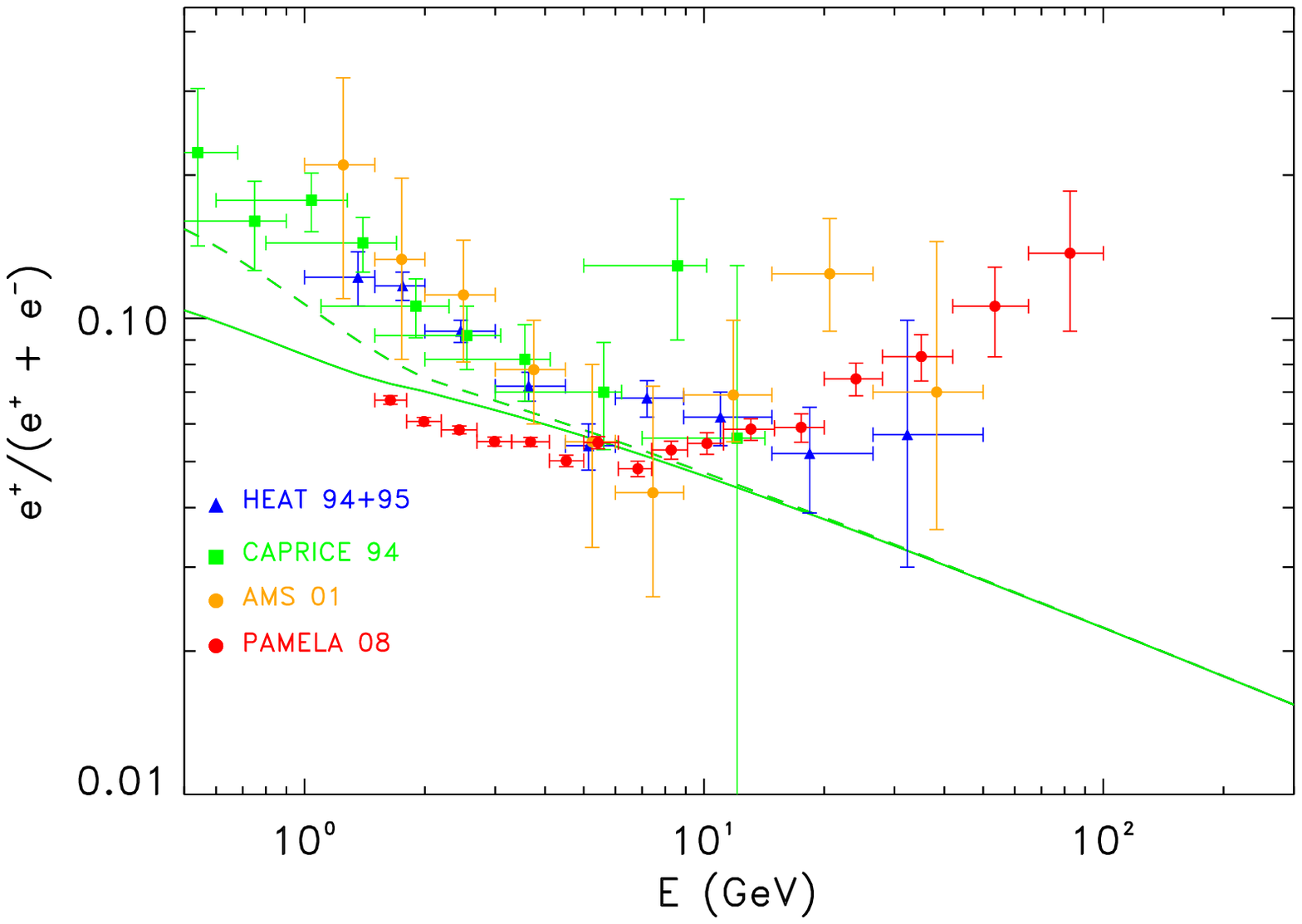}
   \label{fig:pos_PD_singlecomp_best} 
   }
\caption{The electron ($e^- + e^+ $) spectrum is shown for the KOL (panel a), KRA (panel c) and PD (panel e) diffusion setups as specified in Tab.~\ref{table1}.  The electron source spectral indexes are 1.60/2.50 below/above 4 GeV for the KOL model (panel a) and 2.00/2.43 and 2.0/2.400  below/above 2 GeV for the KRA 
(panel c) and PD models (panel e) respectively. Purple and orange lines represent secondary electron and positron spectra respectively.
The corresponding positron fraction ($e^+/(e^- + e^+ )$) curve computed under the same conditions are shown in the panels (b), (d) and (f) respectively.
Solid lines are modulated with a potential $\Phi = 550~{\rm MV}$. Dashed lines are the corresponding LIS spectra. }
\end{figure}

A reasonable fit of the CRE spectrum measured by Fermi-LAT is possible only at the price of normalizing the models to data at 10 GeV (rather than 100 GeV) and adopting  injection spectra slightly steeper than those reported in the above ($2.50$ rather than $2.45$ in the KRA model, $2.43$ rather than $2.37$ for the KOL model). 
However, this is not the most natural choice for the reasons we explained above. Moreover, the observed spectral slope between 7 and 100 GeV is never reproduced in those cases, as it is clear from Fig.~\ref{fig:KOL_singlecomp_best}, \ref{fig:KRA_singlecomp_best} and \ref{fig:PD_singlecomp_best}).  

For the same models, we also computed the corresponding positron fraction $e^+/(e^- + e^+)$ 
(see Fig.~\ref{fig:pos_KOL_singlecomp_best}, \ref{fig:pos_KRA_singlecomp_best} and \ref{fig:pos_PD_singlecomp_best}). 
Clearly, below 10 GeV the $e^+/(e^- + e^+)$  measured by PAMELA can be reproduced by the KRA and PD models while the fit is unsatisfactory for the KOL model. Again, low reacceleration models seem to provide a better description of low energy experimental data. 
None of the single component realizations, however,  can reproduce PAMELA data above 10 GeV. 

We conclude this section by summarizing the main drawbacks of single component CRE models:

\begin{itemize}

\item they are unable to reproduce all the features revealed by Fermi-LAT  in the CRE spectrum, in particular the flattening observed at around 20 GeV (which was also recently found by PAMELA \cite{Adriani_talk}) and the softening at $\sim 500$ GeV. 
If, as they should, such models are normalized against data in an energy range where systematical and theoretical uncertainties are the smallest, they clearly fail to match CRE Fermi-LAT and PAMELA data below 20 GeV.  Most importantly, the CRE spectral slope between 7 and 100 cannot be reproduced.
	
\item They suffer from relevant problems also at very high energy, because they do not reproduce the softening of the CRE spectrum measured by H.E.S.S. above 1 TeV.  This requires either to introduce a cutoff in the CRE source spectrum, which, however, does not account for the observed spectral hardening around 100 GeV, or to introduce an additional CRE component of local origin, with a spectrum peaked around the TeV,  as first suggested in \cite{Aharonian:1995zz}.

\item As explained in many papers (see e.g.~\cite{Serpico:2008te}), they cannot reproduce the rising positron-to-electron ratio measured by PAMELA at high energy. Therefore, if PAMELA observations are correct, an additional positron component, besides that produced by CR spallation, has to be invoked. 

\end{itemize}

\section{Two components models}
\label{sec:toy_model}


In the following we try to reproduce Fermi-LAT, H.E.S.S. and PAMELA CRE and positron data in the framework of a two component scenario. 
The approach we follow in this section is a rather straightforward generalization of what we did in the previous one: we assume the presence in the Galaxy of two CRE source populations rather than one. The first component accelerates only electrons (the standard component) with a spectral slope $\gamma_{e^-} \gtrsim 2$ (as expected from Fermi acceleration in SNRs) up to some cutoff $E^{e^-}_{\rm cut}$, 
while the second one accelerates electrons and positrons with a common spectrum (extra-component) with a form
\begin{equation}
Q_{e^\pm}(E) = Q_{0}\left(\frac{E}{E_{0}}\right)^{- \gamma_{e^\pm}}e^{-E/E^{e^\pm}_{\rm cut}}\;,
\label{eq:extracomp}
\end{equation}
and $\gamma_{e^\pm} <  2$ as required to account for the positron fraction rise observed by PAMELA. 
Here we also assume that both source classes have the same continuous spatial distribution as that considered in the previous section. 
This assumption, however, is not critical, and we checked that a quite wide sample of extra component source distributions (including the one expected from the annihilation of dark matter with a NFW profile) yields quite similar results after a small tuning of the injection parameters. The assumptions of considering the source distribution continuous and cylindrically symmetric up to $\sim 1~\TeV$ are more crucial in this respect and will be reconsidered in the next Sections. 
 
In order to tune the normalization of the two spectral components we follow a multi-messenger approach. Our first step is to tune the standard component to consistently reproduce both the $e^- + e^+$ spectrum measured by Fermi-LAT and the  $e^+/(e^- + e^+)$ measured by PAMELA below 20 GeV, where the effect of the extra component is supposed to be negligible. Remarkably, this is indeed possible if we use propagation setups with low reacceleration, namely either the KRA or the PD setups. 
Noticeably, these setups (especially the KRA one \cite{DiBernardo:2009ku}) also provide the best combined fits of the B/C data, proton and the antiproton spectra measured by PAMELA (see Sec.~\ref{sec:propagation}). Fermi-LAT and PAMELA electron, positron fraction and antiproton data are all reproduced with the same modulation potential $\Phi = 550~\MV$, in the simple force field framework. 
The required source spectral slopes for the electron standard component is  $\gamma_{e^-} = 2.00/2.65, \ 2.00/2.60$  below/above 4 GeV
for the KRA and PD setups respectively. We also introduce an exponential cutoff in the source spectrum of this component at 3 TeV. The choice of higher cutoff would not affect significantly or final results. 

\begin{figure}[tbp]
  \centering 
  \subfigure[]
 {
  \includegraphics[scale=0.4]{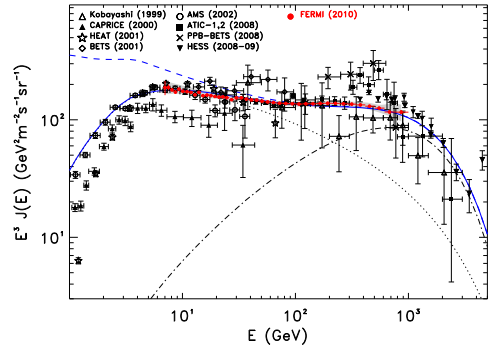}
  \label{fig:extra_component_fit}
  }
   \subfigure[]
 {
\includegraphics[scale=0.4]{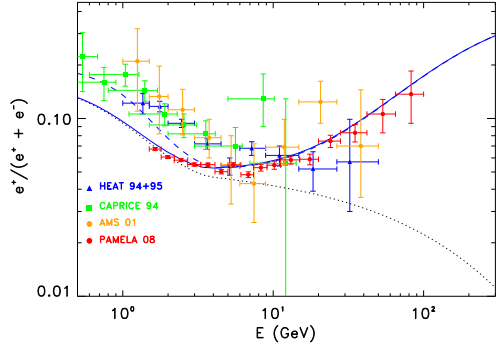}
  \label{fig:extra_component_pamela}
  }
 \caption{ The $e^- + e^+$ total spectrum (panel a) and positron fraction (panel b) data are compared with the predictions of our two-component model. The dotted line represents the propagated standard component with injection slope $\gamma_{e^-} = 2.00/2.65$ above/below $4~\GeV$
 and $E^{e^-}_{\rm cut} = 3~\TeV$, while the dot-dashed line is the $e^\pm$ component with  $\gamma_{e^\pm} =  1.5$ and $E^{e^\pm}_{\rm cut} = 1.4~\TeV$.  Both components are propagated with {\tt DRAGON} adopting the KRA setup. The blue solid/dashed lines represent the modulated/LIS quantities.
 The modulated lines have been computed adopting the charge symmetric modulation potential $\Phi = 550$ MV.}
\end{figure}

Clearly, in the absence of the extra $e^\pm $component, high energy CRE (Fig.~\ref{fig:extra_component_fit}) and positron fraction (Fig.~\ref {fig:extra_component_pamela}) data would completely be missed (see dotted line in Fig.~\ref{fig:extra_component_fit}).
Remarkably, the indication of the presence of an extra CRE component could be found even considering only Fermi-LAT and the low energy PAMELA data alone. 
This a new and interesting result which was made possible by the spectacular data collected by Fermi-LAT and PAMELA and by the fact they operate during the same solar activity phase (hence reducing the uncertainties due to solar modulation).

Our next step is then to suitably choose the extra-component as to reproduce Fermi-LAT and H.E.S.S. high energy CRE data. 
We find here that this is possible by taking $\gamma_{e^\pm} = 1.5$ and  $E_{\rm cut} = 1.0 \div 1.5 ~\TeV$ both for the KRA setup (see Fig.~\ref{fig:extra_component_fit}) 
and the PD one.
This is similar to what done in \cite{CRE_interpretation1,DiBernardo:2009iu} but for the choice of the propagation setup which in those papers was assumed to obey Kolmogorov diffusion (as a consequence, low energy PAMELA data were not reproduced in that case). 

It is interesting that preliminary $e^-$ spectrum measured by PAMELA \cite{Adriani_talk} is also nicely reproduced by our extra-component models. Above 100 GeV this spectrum is softer than the $e^- + e^+$ measured by Fermi-LAT by the exact amount which is required to leave room the $e^+$ extra-component. 
It is remarkable that such a relatively simple approach allows to reproduce such a large number of observations.

The fact that the high energy tail of the CRE spectrum is reproduced so well in terms of a continuous extra component source distribution suggest that either it is dominated by a physically continuous distribution of sources (as one expects for dark matter) or by a single astrophysical source in the nearby. 
In the next section we will discuss the latter case.

\section{The role of astrophysical nearby sources}\label{sec:discrete_sources}


The nature of the extra-component of primary electrons and positrons that we invoked in the previous section is an intriguing matter of debate, and the possible scenarios include both an exotic explanation (involving annihilation or decay of Particle Dark Matter) or a purely astrophysical interpretation. Here we concentrate on the second possibility and show that well known astrophysical sources can rather naturally account for the available CRE electron and positron fraction observations.  
In Sec.~\ref{sec:anisotropy} we will show that CRE anisotropy measurements should soon be able to confirm this possibility.

Differently from the previous section, here we treat the extra component as originating from a discrete collection of sources rather than from a continuous distribution. This is the proper approach to be followed for energies exceeding few hundreds GeV at which the loss length due to synchrotron emission and IC scattering becomes comparable to the average SNR mutual distance so that only few sources within few hundred parsecs are expected to dominate the CRE flux.
To be even more realistic, we work with actually observed astrophysical CRE source candidates,  pulsars and SNRs, as taken from radio catalogues.  We then treat electron and positron propagation from those source to the Solar system by solving analytically the diffusion-loss equation similarly to what done in \cite{CRE_interpretation1}. Since the analytical expressions and tools used here are the same followed in that work we do not report here any detail and address the interested reader to that paper.  

Concerning the large scale (standard) component, we model it with {\tt DRAGON} as done in the previous section. 
For consistency, we treat analytical and numerical propagation under the same physical conditions.        

Since the extra component does not affect the low energy tail of the CRE and positron spectra, all we did in the previous section to fix the standard component holds also here. 
Therefore we again assume $\gamma_{e^-} = 2.65$ and a KRA propagation setup. Similar results can be obtained with a PD setup.  This is different from what done in \cite{CRE_interpretation1} where the KOL setup was used.
   
\subsection{The contribution of local pulsars}\label{sec:pulsars_only}

Pulsars are extreme astrophysical environments that release very large amounts of energy ($\sim 10^{52} \div 10^{54}$ erg) during their active lifetime. Most of this energy is emitted in the early stage of their evolution, since they are powered by their spin-down.  They are believed to be sources of both electrons and positrons (see e.g.~\cite{Shen,Harding:87}): in fact, magnetospheric models predict the formation of curvature photons that -- interacting with the strong pulsar magnetic field -- form electron-positron pairs.  Those particles are then expected to be accelerated at the termination shock of the pulsar wind nebula (PWN) and be released in the ISM when the PWN merges the ISM or the pulsar exit the parent SNR shell due to its proper motion (see \cite{Blasi:2010de}  for a detailed description of the latter scenario).  In both cases, this process typically happens in a time between $10^4$ and $10^5$ years.

The possibility that electrons and positrons from nearby pulsars can dominate the high energy tail of the CRE spectrum and explain the rising behavior of the positron fraction was already proposed in \cite{Aharonian:1995zz} and studied in several more recent papers (see e.g.~\cite{Hooper:2008kg,Profumo:2008ms}). 
 
Similarly to the approach taken in \cite{CRE_interpretation1}, we model the CRE emission from pulsars as a point-like burst with a time delay with respect to the birth of the object. Such a delay is required to reproduce the smooth CRE spectrum measured by Fermi-LAT and H.E.S.S. and is motivated by the fact that electrons and positrons are expected to be trapped in the nebula until it merges with the ISM. We assume that electrons and positrons are accelerated in equal amounts by pulsars with a power-law injection energy up to an exponential cutoff  at energy around 1 TeV.  

We considered the pulsars within 2 kpc from Earth, taken from the ATNF catalogue \cite{Manchester:05}
\footnote{http://www.atnf.csiro.au/research/pulsar/psrcat/}. 
We verified that more distant pulsars give a negligible contribution. We also verified that $\gamma$-ray pulsars which have been detected by Fermi-LAT \cite{blind_search}
and are not in the ATNF catalogue, which can also contribute to the observed $e^- + e^+$ spectrum (see \cite{Gendelev:2010fd}) do not affect significantly our results. 

For middle-aged pulsars, the rotational energy released at the time T can be approximated as $\displaystyle E(T)  \simeq   {\dot  E}_{\rm PSD}~\frac{T^2}{\tau_0}~$, where ${\dot  E}_{\rm PSD}$ is the spin-down luminosity determined from the observed pulsar timing and $\tau_0$ is a characteristic braking time. Due to the strong energy losses in the PWN,  only electron-positron pairs which are accelerated before the escape time $T_{\rm esc}$ can contribute to the CRE flux. Therefore the $e^\pm$ energy released in ISM is  
$\displaystyle E_{e^\pm}(T)  \simeq \eta_{e^\pm}~  {\dot  E}_{\rm PSD}~\frac{\left(T - T_{\rm esc}\right)^2}{\tau_0}~$, where $\eta_{e^\pm}$ is the $e^\pm$ pair conversion efficiency. 

\begin{figure}[tbp]
 \centering
  \subfigure[]
 {
  \includegraphics[scale=0.4]{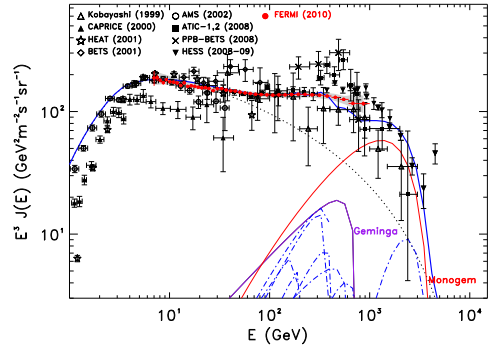}
  \label{fig:pulsars}
  }
   \subfigure[]
 {
\includegraphics[scale=0.4]{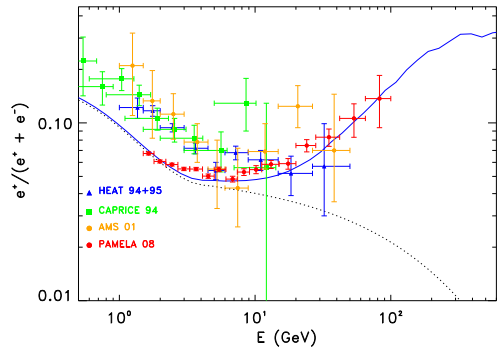}
  \label{fig:pulsars_pamela}
  }
\caption{The contribution from nearby pulsars (within 2 kpc) computed analytically for the KRA diffusion setup is added to a standard component
similar to that shown in Fig.~\ref{fig:extra_component_fit} but the addition of an exponential cutoff at 3 TeV. Panel a): electron+positron spectrum. Panel b): positron fraction. Notice that since pulsars are assumed to emit both electrons and positrons, the rising positron-to-electron ratio measured by PAMELA is correctly reproduced within such a scenario.  The pulsar $e^\pm$ conversion efficiency is $\simeq 35\%$}
\end{figure}

As shown in Fig.~\ref{fig:pulsars} and \ref{fig:pulsars_pamela}, the pulsar scenario allows a very good fit of both high-energy Fermi-LAT and H.E.S.S. electron+positron as well as PAMELA positron fraction data, due to the presence of primary positrons. The plots in Fig.s \ref{fig:pulsars} and \ref{fig:pulsars_pamela} have been computed assuming that  pulsars spin down as a magnetic dipole (braking index $n = 3$) and that $\tau_0 = 10^4~\yr$,  as commonly done in the literature. Furthermore we tuned the pulsar injection parameters to the values $\gamma_{e^\pm} = 1.4$, $E_{\rm cut} = 2~\TeV$, $T_{\rm esc} = 75~\kyr$ and $\eta_{e^\pm} = 35\%$. These values are compatible with multichannel observations of pulsars (see  e.g.~\cite{Hooper:2008kg,Blasi:2010de} and ref.s therein).  Significantly larger spin-down power, hence smaller values of $\eta_{e^\pm}$, are obtained for values of the braking index smaller than 3 as observed for several pulsars (note e.g.~that for the Crab pulsar $n = 2.5$), which make this scenario even more palatable.

While it is certainly unrealistic to assume that all pulsars share the same values of those parameters, this is not critical for our results since the high energy tail of the spectrum is always dominated by a single object, namely the Monogem pulsar (PSR B0656+14).  This is a consequence of its small distance  ($d \simeq 288^{+ 33}_{- 27}$ pc) \cite{Brisken:2003hs}, relatively young age ($T \simeq 1.1 \times 10^5\, \yr $) and spin-down luminosity ${\dot E} \simeq 3.8 \times 10^{34}~\erg \s^{-1}$. 
Monogem available rotational energy at the adopted $e^\pm$ escape time is  $E(t > T_{\rm esc}) \simeq 5 \times 10^{47}\, {\rm erg}$.   
Furthermore, it was shown in \cite{CRE_interpretation1} that it is possible to reproduce CRE and positron data for several allowed combinations even randomly varying the pulsar parameters. 

Due to the time delay between their birth and the $e^\pm$ release, very young pulsars, such as Vela, which are bright in the GeV and TeV gamma-ray sky, do not contribute, which explains the absence of  pronounced bumps in the CRE spectrum around or above the TeV which were instead predicted in several other papers (see e.g.~\cite{Kobayashi:2003kp,Profumo:2008ms}).  

\subsection{Adding the contribution from local SNRs}\label{sec:hydric_model}

Similar to the approach of \cite{CRE_interpretation1},  in the above we treated the standard electron component as originated by a continuous distribution of sources.  This approximation, however, is not realistic above few hundred GeV for the reasons we explained at the beginning of this Section.  Indeed, if we share the common, and well motivated,  wisdom that CRE are accelerated by SNR, only few of those objects will contribute to the CRE observed spectrum above 100 GeV.  This may produce observable features in the total CRE spectrum.

In order to study such effect, we treat CRE propagation from nearby SNRs similarly to what we just did for pulsars. Since the SNR lifetime is typically smaller than the propagation time, we consider the emission from a single SNR  as a single burst simultaneous to the SNR birth. 
Hence, we consider all observed SNRs within 2 kpc as taken from the Green catalogue \cite{Green:2009qf} and treat them as point-like $e^-$ sources with a power-law injection spectrum and an exponential cutoff. 

\begin{figure}[tbp]
\begin{center}
 \centering
  \subfigure[]
 {
  \includegraphics[scale=0.4]{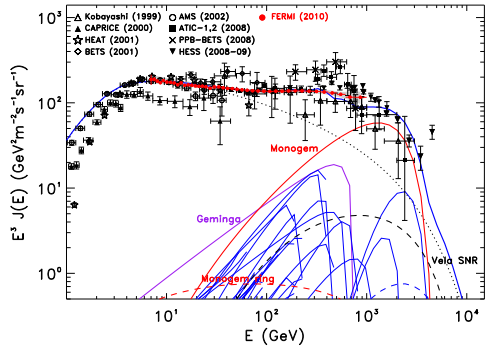}
  \label{fig:Hybrid_1_spectrum}
  }
  \subfigure[]
 {
 \includegraphics[scale=0.4]{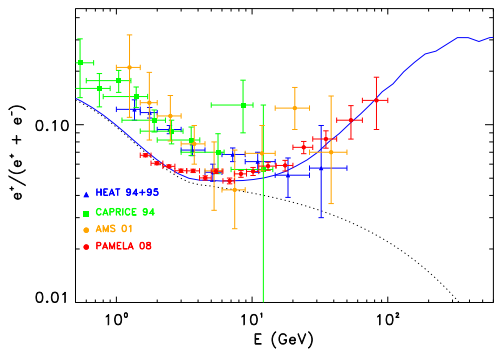}
  \label{fig:Hybrid_1_pamela}
  }
\caption{The analytically computed CRE flux from both nearby (within 2 kpc) SNRs and pulsars is added to the same standard component used in Fig. \ref{fig:pulsars}.  Again, all components are propagated using the KRA diffusion-reacceleration setup.
Panel a): electron+positron spectrum. Panel b): positron fraction. The assumed energy release for each SNR is taken as $2 \times 10^{47}\,{\rm erg}$ . The pulsar efficiency is $\simeq 30\%$. Solar modulation modulation potential is $\Phi = 500 MV$. }
\end{center}
\end{figure}

In Fig.s \ref{fig:Hybrid_1_spectrum} and \ref{fig:Hybrid_1_pamela}  we respectively represent the CRE spectrum and positron fraction obtained for a reasonable combination of pulsar and SNR parameters, namely:  spectral index $\gamma_{e^{- \,{\rm SNR}}} = 2.4$;  cutoff energy $E^{\rm SNR}_{\rm cut} = 2~\TeV$;  electron energy release per SN  $E^{\rm SNR} =  2 \times 10^{47}\,{\rm erg}$;   $\eta_{e^\pm} \simeq 30\%$ (which is slightly smaller than that needed without considering nearby SNR) for all pulsars. We see from Fig. \ref {fig:Hybrid_1_spectrum}  that under those conditions, the dominant source in the TeV region remains Monogem pulsars.

Clearly, other combinations of parameters are possible, and the relative contributions of the several sources may vary.   
However, the requirement to reproduce the PAMELA positron fraction imposes an important independent constraint which does not permit to lower significantly the Monogem dominant contribution with respect to that of SNRs.  
Therefore the discrete contribution of nearby SNRs should not introduce pronounced features (bumpiness) in the CRE spectrum.
Another relevant, and testable, prediction of this scenario is that, being the high energy CRE flux  dominated by Monogem pulsar, a significant CRE dipole anisotropy should be present roughly directed towards that pulsar.
This possibility will be discussed in details in the next Section.  

\section{The CRE anisotropy}\label{sec:anisotropy}


Recently the Fermi-LAT Collaboration published upper limits on the anisotropy in electron + positron flux \cite{Ackermann:2010ip}. It is therefore worth checking if the models discussed in the previous section are compatible with such limits.

The analytical expressions which we used to compute the CRE anisotropy due to a single pulsar is:
\begin{equation}
{\rm Anisotropy}  =  \frac{3}{2 c} \frac{r}{T-T_{\rm esc}}~\left( \frac{1 - (1 - E/E_{max}(t))^{1 - \delta} } {(1 - \delta)E/E_{max}(t) } \right)^{-1} \; \frac{N_e^{\rm PSR}(E)}{N_e^{\rm tot}(E)}
\end{equation}
where $N_e^{\rm PSR}$ and $N_e^{\rm tot}$ are the electron spectra from the pulsar and its sum to the large scale Galactic plus distant pulsar components; $T$ and $T_{\rm esc}$ are the pulsar birth time and the time it takes for the electrons to be released in the ISM respectively; $E_{\max}(t)$ is the energy loss time due to synchrotron and IC losses (see \cite{CRE_interpretation1,Kobayashi:2003kp} for more details).

\begin{figure}[tbp]
 \centering
  \subfigure[]
 {
 \includegraphics[scale=0.4]{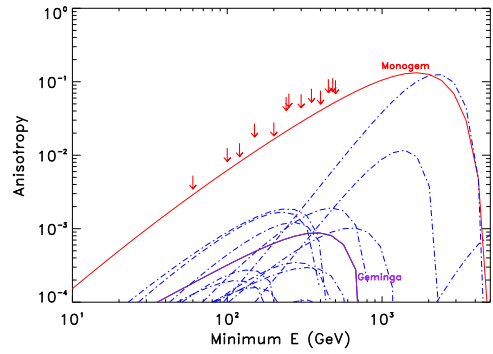}
  \label{fig:pulsars_aniso_new}
  }

   \subfigure[]
 {
 \includegraphics[scale=0.4]{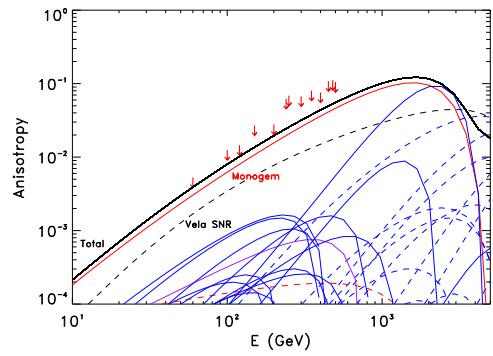}
  \label{fig:Hybrid_1_aniso}
  }
\caption{Panel a): 
the integrated anisotropy, as a function of minimum energy,  computed for the pulsar model plotted in Fig.~\ref{fig:pulsars}
is compared to the 95 \% C.L. Fermi-LAT upper limits \cite{Ackermann:2010ip}; 
Panel b) The same is done here for the pulsars + SNRs model plotted in Fig.~\ref{fig:Hybrid_1_spectrum}.
The black solid line represents the total anisotropy, originated mostly by Monogem and Vela SNR. 
}
\end{figure}

The most important results we obtained are summarized in Fig.~\ref{fig:pulsars_aniso_new} - \ref{fig:Hybrid_1_aniso}.

First of all, it is important to notice that the model discussed in Sec.~\ref{sec:pulsars_only}  (see Fig.~\ref{fig:pulsars}), where only the emission from nearby pulsars is added to the smooth Galactic standard component, is compatible with the upper limits reported by the Fermi-LAT Collaboration. The middle-aged pulsar Monogem gives the dominant contribution to the anisotropy, and the expected value of this observable is very close to the upper limit, so that a positive detection is expected in the near future towards Monogem.

Hence we now consider the model discussed in Sec.~\ref{sec:hydric_model} (see Fig.~\ref{fig:Hybrid_1_spectrum}), in which the main contribution to the high energy $e^- + e^+$ flux still comes from pulsars but the contribution of local SNRs is also considered (hybrid model). 
We see from Fig.~\ref{fig:Hybrid_1_aniso} that also this possibility is not excluded by anisotropy measurements: in this plot, the reader may notice that the Monogem pulsar (red solid line) and the Vela SNR (black dashed line) contribute most to the total anisotropy (the black solid line), which is computed as the sum of each anisotropy weighted by the cosine of the angle of the corresponding source with respect to the direction of the maximum flux. However, also in this case the total expected anisotropy is very close to the measured upper limit, so that a future detection at level $\sim 1\%$ at $\sim 1$ TeV towards the portion of the sky where Vela and Monogem are located (with the peak situated almost in the middle between the two sources) is to be expected in the next years.

\begin{figure}[tbp]
 \centering
   \subfigure[]
 {
 \includegraphics[scale=0.4]{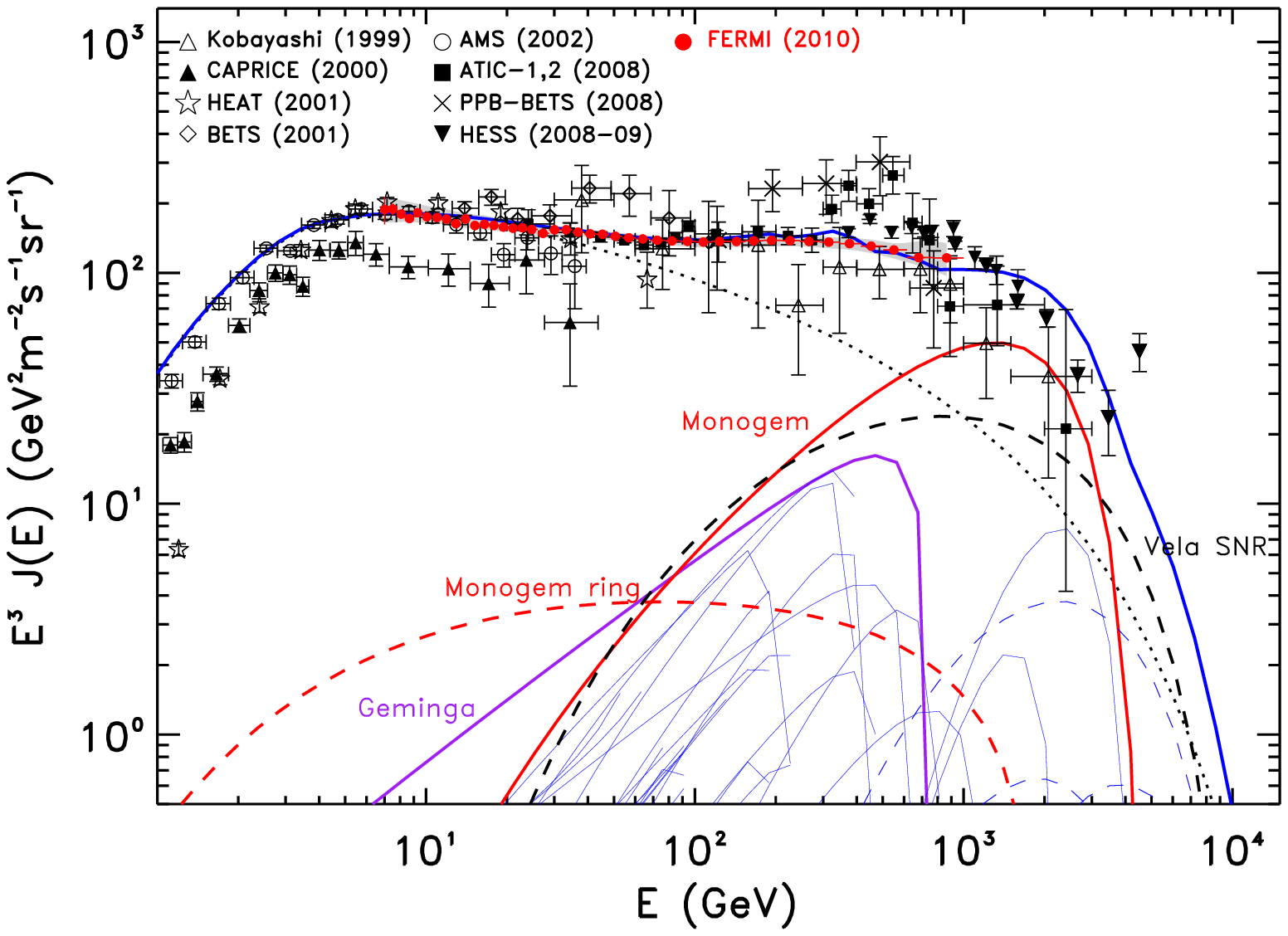}
  \label{fig:Hybrid_2_spectrum}
  }
   \subfigure[]
 {
 \includegraphics[scale=0.4]{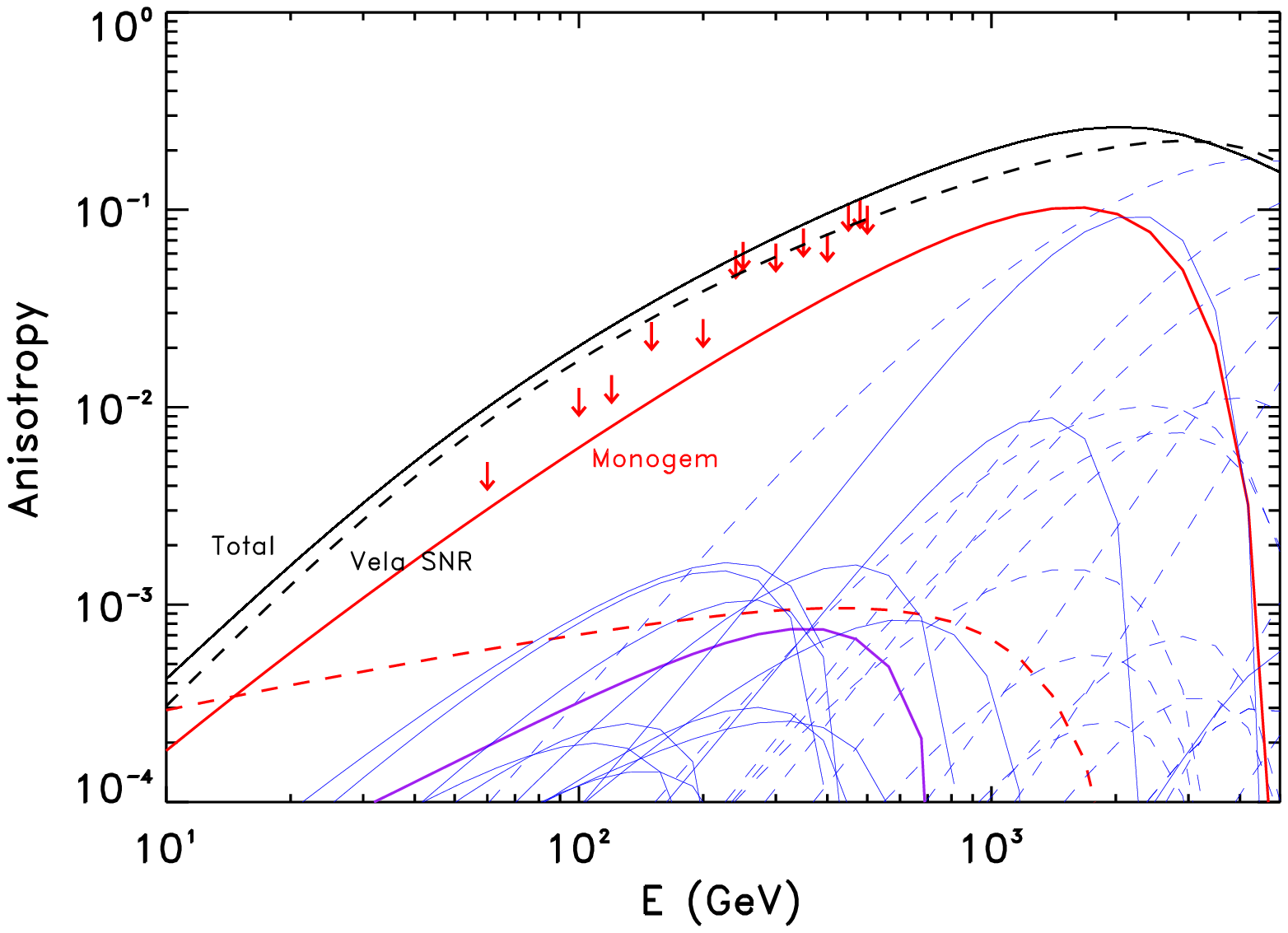}
  \label{fig:Hybrid_2_aniso}
  }
\caption{Panel a): 
The CRE spectrum for a modified version of the pulsars + SNRs (hybrid) model in which the energy output of SNRs is increased to $1 \times 10^{48}\,{\rm erg}$ is represented.  This model is still compatible with Fermi-LAT and HESS electron data. Panel b) The corresponding anisotropy is shown: a strong contribution from SNRs is not compatible with Fermi-LAT upper limits.}
\end{figure}

It should be noticed that Fermi-LAT anisotropy constraints already exclude some models which do reproduce the CRE and the positron fraction data. 
For illustrative purposes, in Fig.s \ref{fig:Hybrid_2_spectrum} and \ref{fig:Hybrid_2_aniso}  we show the CRE spectrum and anisotropy computed for a
 a model in which the electron energy emitted by  SNRs is increased from $2 \times 10^{47}\,{\rm erg}$ to $1 \times 10^{48}\,{\rm erg}$ and
pulsar $e^\pm$ emission efficiency is consequently reduced from $35 \%$ to $30 \%$ so to reproduce the observed CRE spectrum.
We see  from Fig.~\ref{fig:Hybrid_2_aniso} that the expected anisotropy exceeds the upper limits in that case, mostly because of the large contribution of Vela SNR, which is nearby (less than 300 pc) and very young ($\sim 10^4 \, {\rm yr}$). Although this result can hardly be translated into a sharp constraint on the $e^\pm$ energy output from SNRs,  due to the large number of free parameters we had to deal with,  nevertheless we can safely conclude that a scenario in which SNRs provide the dominant contribution to CRE spectrum in the TeV region is incompatible with Fermi-LAT upper limits on the CRE anisotropy. 

\section{Discussion}

Here we discuss how general are the physical assumptions under which we reproduce such a large number of CR data sets and the compatibility of our results with other complementary observations. 

We start discussing the source spectrum of the electron standard component. According to the common wisdom we assumed that this component is accelerated by Galactic SNRs. The most clear evidence that electron acceleration takes place in SNRs comes from the synchrotron emission of those objects. The mean value of the spectral index from the Green's catalogue \cite{Green:2009qf} is  $0.5 \pm 0.15$  in the GHz range, which loosely implies $\gamma = 2.0 \pm 0.3$ in the 1 - 10 GeV range \cite{Delahaye:2010ji}. This is therefore compatible with the source spectral index adopted in Sec.s  \ref{sec:toy_model} and \ref{sec:discrete_sources} below few GeV. At larger energies, where we need $\gamma_{e^-} \simeq 2.6$, we have less stringent observational constraints. 
While $\gamma$-ray measurements favor harder spectral indexes above 10 GeV, it is not clear if the origin of that emission is leptonic. Recent $\gamma$-ray observations of SNRs actually are better described in terms of a hadronic origin of that radiation (see e.g.~\cite{Abdo:2009zz,Abdo:2009gg}).  On the theoretical side, Fermi linear acceleration in SNRs generally predicts values of the spectral index smaller than 2.5, which is marginally at odds with our finding. However, linear Fermi acceleration theory is likely to be not exact and the actual spectral shape could be different from that prediction. 
Furthermore, it should be noted that we modeled the standard component in the approximation of a cylindrically symmetric source distribution. Such description is adequate to reproduce nuclei spectra and secondary-to-primary ratios, but may be less realistic for electrons in the 0.1 - 1 TeV range where the local distribution becomes relevant.  A more realistic distribution which accounts for the spiral arm distribution of SNRs may actually result in a different requirement for the injection spectral slope fitting the data in that energy range. Indeed, being the Sun at the edge of a Galactic arm, the average distance from SNR is larger than in the smooth case. As a consequence, a harder injection spectrum is required in this case to compensate for the larger energy losses and reproduce the observed spectrum.
We performed explorative runs probing the effects of a Galaxy spiral structure which actually confirm this expectation without spoiling the rest of our successful results. A more detailed study of this effect will be performed elsewhere.

For what concerns the $e^\pm$ extra-component, we mentioned as the hard spectrum ($\gamma_{e^\pm}  \simeq 1.5$) required to explain experimental data
can be originated by pulsars wind nebulae (PWNe).  While the acceleration mechanism responsible for such spectral shape is not understood yet,  
on purely observational grounds we know that PWNe indeed accelerate electrons with a power law spectrum which is as flat as $E^{-1} - E^{-1.8}$  up to several hundred GeV \cite{Blasi:2010de}.   
The most critical issue here concerns the available energy that middle-age pulsars can release under the form of $e^\pm$ pairs.  
We showed, however, that under reasonable conditions the rotational energy is sufficient to account for experimental data even if pairs are released in the ISM after several $10^4~\yr$.  A physically viable scenario where $e^\pm$ are accelerated in bow shock PWNe has been recently discussed in \cite{Blasi:2010de}. The injection parameters adopted here are fully compatible with those proposed in that work. Forthcoming Fermi-LAT CRE spectrum and anisotropy measurements as well as $\gamma$-ray observations of PWNe may help to validate this scenario in the next future. 

Since CR electrons contribute to the $\gamma$-ray diffuse emission of the Galaxy by their bremsstrahlung and IC radiation, it is worth asking if the two component models presented above are compatible with Fermi-LAT observations also in that channel. For this purpose, we used our numerical package {\tt GAMMASKY} and cross checked our results with {\tt GALPROP}.  For this purpose we used the same radiation field as described in \cite{Porter:2006tb}.
To derive a detailed model of the $\gamma$-ray diffuse emission of the Galaxy is far beyond the aims of this paper.  
We only computed the $\gamma$-ray spectrum at intermediate Galactic latitudes $10^{\circ} < \vert b \vert < 20^{\circ}$, averaged over all longitudes, and compared it with Fermi-LAT data presented in \cite{Abdo:2009mr}. For each CR diffusion setup we computed the $\pi^0$ component consistently.
We treated the standard and the extra-component CRE components as described in Sec.~\ref{sec:toy_model}.  The contribution of local CRE sources is almost irrelevant here since the $\gamma$-ray flux is dominated by distant sources. 
We found that all models discussed in that sections reasonably reproduce the $\gamma$-ray spectrum measured by Fermi-LAT and within errors agree with the model presented in \cite{Abdo:2009mr}.  We also found that the contribution of the extra-component electron to the bremsstrahlung and IC  $\gamma$-ray spectrum is subdominant and may be hard to detect.

\section{Conclusions}

The recent extension of the measurement of the CRE spectrum down to 7 GeV by the Fermi-LAT collaboration offers a strong evidence of the presence of a new spectral component.   In fact,  even allowing for a spectral break in the CRE source spectrum at few GeV,  single component models fail to provide a satisfactory fit of those data.  Such evidence adds to the one provided by PAMELA observations of a rising behavior of the positron fraction above 10 GeV. 
Assuming that the new component is symmetric in electron and positrons,  we showed that a consistent interpretation of a wide sets of data is possible. For the first time, this also includes PAMELA positron fraction data even below 10 GeV as well as proton and antiproton data taken by the same experiment which are crucial to constrain the allowed propagation setup(s) and the solar modulation potential (during the same solar cycle phase Fermi-LAT is operating).  We showed that only few among the propagation setups matching the B/C data can also consistently match CRE and positron fraction data. 
 This is the case for Kraichnan-like and plain-diffusion setups while the commonly adopted Kolmogorov-type setup is disfavored.  
We obtained these results working in a simple force free, charge symmetric, solar modulation framework. This does not means that
charge dependent effects, which are expected to be present at some level due to the complex structure of the solar magnetosphere, are absent but only that they seems not to be required for a consistent interpretation of Fermi-LAT and PAMELA data above few GeV.

Concerning the origin of the $e^\pm$ extra-component, we confirm that observed nearby pulsars are realistic source candidates.
The expected anisotropy in the direction of the most prominent CRE candidate source, Monogem pulsar, is compatible with the present upper limit just released by Fermi-LAT collaboration and may be detectable in a few years. 
We also considered here the effect of known nearby SNRs on the CRE spectrum and the anisotropy, showing that they should not affect significantly those quantities.

We conclude by observing that the AMS-02 space mission \cite{ams02} to be installed on the International Space Station next year will be crucial to validate the models discussed in this paper and to look for small spectral anomalies which may indicate the presence of new physics. 
Beside providing and independent check of the positron anomaly detected by PAMELA and measure it up to 500 GeV or above, AMS-02 simultaneous measurements of CR  nuclei, antiproton, electron and positron absolute spectra will allow to reduce the still large uncertainties on the propagation setup and the effects of solar modulation.  
Furthermore, its good energy resolution - which will be significantly better than Fermi-LAT - may allow to confirm the spectral features hints found in the Fermi-LAT CRE spectrum and to single out yet unidentified ones associated with nearby astrophysical sources or the dark matter.  

\section*{Acknowledgments}
We are indebted with A.W. Strong and A. Moiseev for valuable discussions and suggestions.  We also thank R. Bellazzini, P. Blasi, D. Horns, L. Latronico, G. Sigl for reading the draft and/or providing us useful comments.
We warmly thank P.~Picozza for allowing us to extract preliminary PAMELA proton data from his talk \cite{PAMELA:proton}.

D.~Grasso has been supported by the Italian Space Agency under the contract AMS-02.ASI/AMS-02 n.~I/035/07/0.  
L. M. acknowledges support from the State of Hamburg, through the Collaborative Research program ÔÔConnecting Particles with the CosmosÕÕ within the framework of the LandesExzellenzInitiative (LEXI).

\end{document}